\documentclass{article}
\usepackage{spconf,amsmath,epsfig}
\usepackage{booktabs} 
\usepackage{subcaption}
\usepackage{multirow}
\usepackage{times}
\usepackage{epsfig}
\usepackage{graphicx}
\usepackage{amsmath}
\usepackage{amssymb}
\usepackage{xcolor}
\usepackage{algpseudocode}
\usepackage{algorithm,tabularx}
\usepackage{helvet}  
\usepackage{courier}  
\usepackage{url}  
\usepackage{epstopdf}
\usepackage{amscd,epsfig,amsfonts,rotating}

\newcolumntype{C}[1]{>{\centering\arraybackslash}p{#1}}
\graphicspath{{fig/}}

\begin{document}\sloppy

\def\x{{\mathbf x}}
\def\L{{\cal L}}

\title{Adversarial Cross-Modal Retrieval via Learning and Transferring Single-Modal Similarities}
%
\name{Xin Wen\textsuperscript{1}, Zhizhong Han\textsuperscript{1,2}, Xinyu Yin\textsuperscript{1}, Yu-Shen Liu\textsuperscript{1*}\thanks{*Corresponding author: Yu-Shen Liu}}
\address{
\textsuperscript{1}School of Software, Tsinghua University, Beijing 100084, China \\
Beijing National Research Center for Information Science and Technology (BNRist)\\
\textsuperscript{2}Department of Computer Science, University of Maryland, College Park, USA\\
x-wen16@mails.tsinghua.edu.cn,
h312h@umd.edu,
yinxy15@mails.tsinghua.edu.cn,\\
liuyushen@tsinghua.edu.cn
}
\maketitle

\begin{abstract}
Cross-modal retrieval aims to retrieve relevant data across different modalities (e.g., texts vs. images). The common strategy is to apply element-wise constraints between manually labeled pair-wise items to guide the generators to learn the semantic relationships between the modalities, so that the similar items can be projected close to each other in the common representation subspace.
However, such constraints often fail to preserve the semantic structure between unpaired but semantically similar items (e.g. the unpaired items with the same class label are more similar than items with different labels). To address the above problem, we propose a novel cross-modal similarity transferring (CMST) method to learn and preserve the semantic relationships between unpaired items in an unsupervised way.
The key idea is to learn the quantitative similarities in single-modal representation subspace, and then transfer them to the common representation subspace to establish the semantic relationships between unpaired items across modalities.
Experiments show that our method outperforms the state-of-the-art approaches both in the class-based and pair-based retrieval tasks.
\end{abstract}
\section{Introduction}
\begin{figure}[t]
\includegraphics[width=\linewidth]{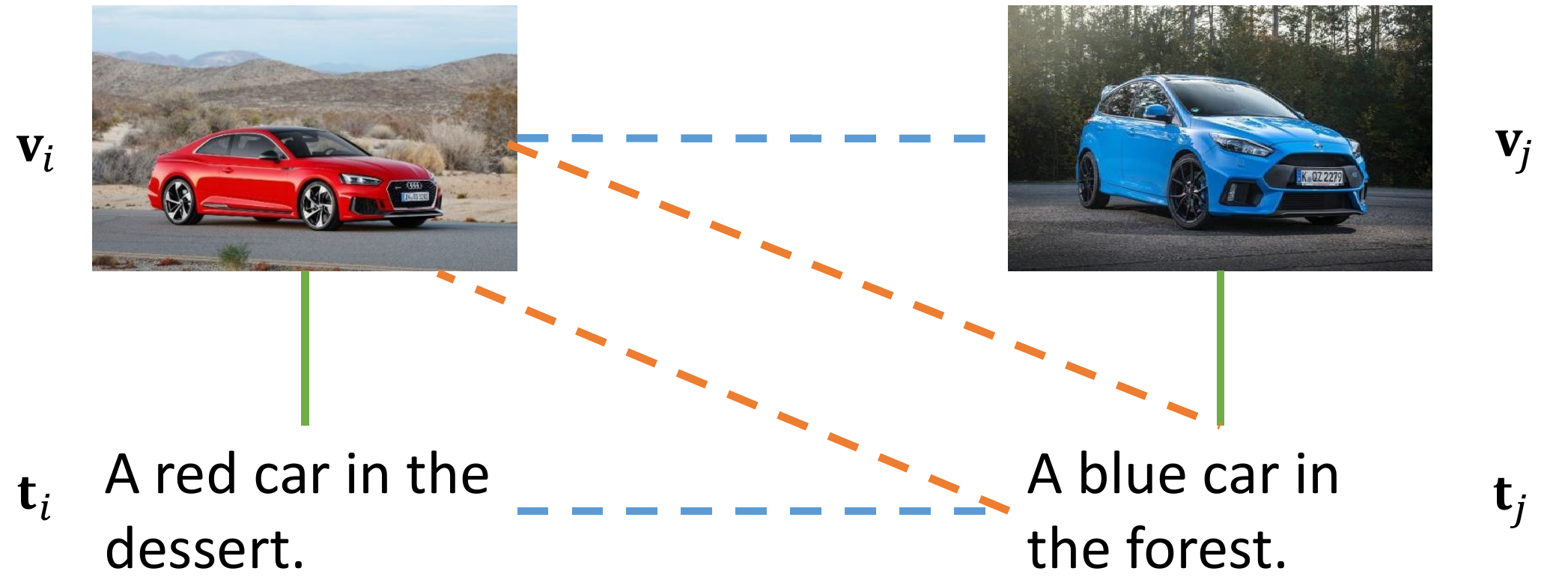}
\caption{Demonstration of the missing semantic relationships. The green solid line is the manual label of paired items cross modalities, and the dot line is the missing label for inter-modal items (orange) and inner-modal items (blue). The proposed CMST considers learning the missing inner-modal similarity (blue dot line) and transferring it to the inter-modal similarity (orange dot line) based on paired items (green solid line).  }
\label{fig:2}
\end{figure}

Cross-modal retrieval aims to retrieve relevant data across different modalities, which enables flexible search across multiple modalities.
The common strategy is to apply element-wise constraints on manually labeled cross-modal pairs to bridge the semantic gaps between different modalities.
However, such manually labeled pairs can only reflect a small part of the semantic structure in the common representation subspace, while the abundant semantic relationships between unpaired items are often failed to be preserved. As demonstrated in Figure \ref{fig:2}, only the similarity between the red car and its corresponding description are manually labeled (green solid line), as well as the similarity between the blue car and its description. However, the similarity between the red car and the blue car's description are missing (orange dot line), although they are semantically similar to some extent because of the same label (the label of car) they share.

To solve the above problem, methods like DCML \cite{liong2017deep} and ACMR \cite{wang2017adversarial} try to utilize the intra-class and inter-class labels of the cross-modal items \cite{mignon2012cmml, wang2017adversarial} by directly assigning the highest similarity (e.g., the value of 1) to the items with the same class label and the lowest similarity (e.g., the value of 0) to the items with different class labels. The problem is, these labels cannot quantitatively reflect the semantic relationships between the intra-class items, which is especially important for the retrieval tasks, because the rank list should indicate the discriminative order of all the retrieved items, especially for the items in the same class. The samples that matches the query better should be ranked higher compared to other samples even in the same class. 
On the other hand, using the intra-class and inter-class labels as the supervision information also makes the cross-modal retrieval methods sensitive to dataset noises such as mislabeled samples. 


In this paper, to address the above-mentioned issues, a novel cross-modal similarity transferring (CMST) method is proposed for cross-modal retrieval. 
The main idea is to employ unsupervised strategy to learn the endogenous semantic relationships between unpaired items in each single-modal representation subspace, and then, transfer the learned relationships to the common representation subspace to establish the semantic structure between unpaired cross-modal items.
In detail, the CMST first employs a similarity learning network to establish the finer similarity metric to capture the semantic structure of training items in each single-modal space. Then, three similarity transferring approaches are proposed to transfer the learnt single-modal similarities to the common representation subspace. CMST works in an adversarial framework to utilize the ability of distribution generation from generative adversarial networks. Our main contributions can be summarized as follows.

\begin{itemize}
\item A novel CMST method is proposed to learn and preserve the semantic structure between unpaired items across different modalities in the common representation subspace in an unsupervised way.
\item Three similarity transferring approaches are proposed by the observation on how cross-modal relationship is built from the similarities in single modalities, which are proven effective in the experiments.
\end{itemize}

\begin{figure*}
\centering
\includegraphics[width=\textwidth]{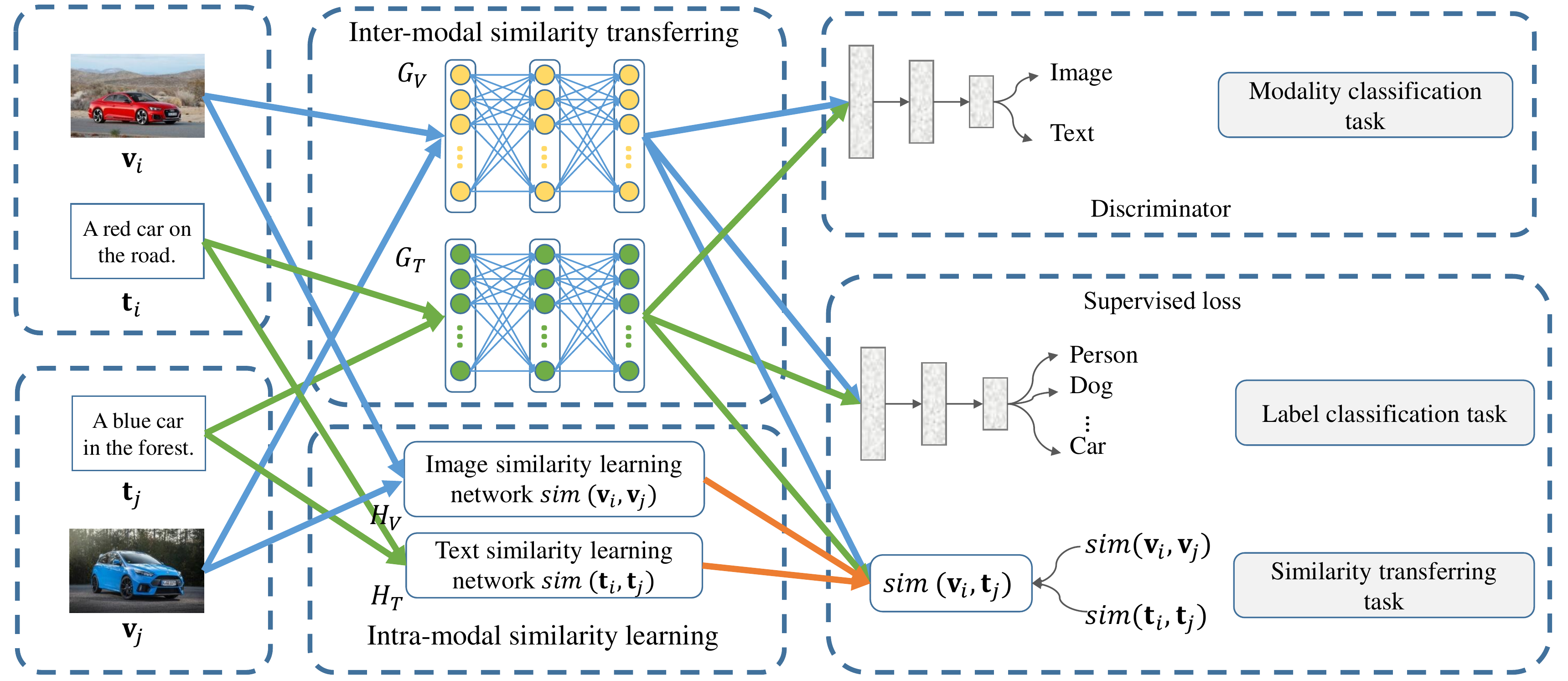}
\caption{The overall structure of the CMST method, including two intra-modal similarity learning networks and an inter-modal similarity transferring network. This figure shows the procedure to learn the cross-modal similarity between $\mathbf{v}_i$ and $\mathbf{t}_j$. Different from the existing methods which simply assign the highest similarity to intra-class cross-modal samples, the CMST method uses the similarity value between $\mathbf{v}_i$ and $\mathbf{v}_j$ to guide cross-modal similarity learning, which is learnt by the intra-modal similarity learning network in the image modality.}
\label{fig:0}
\end{figure*}
\section{Related Work}
Cross-modal retrieval methods can be roughly divided into \textit{joint representation learning methods} \cite{wang2017adversarial} and \textit{cross-modal hashing methods} \cite{song2018binary, xu2017learning}. The proposed CMST model falls into the category of joint representation based methods. It aims to learn a real common representation subspace of multimodal data, where cross-modal data can be directly compared to each other through predefined similarity measurement. Cross-modal retrieval methods like CCA-based methods \cite{andrew2013deep, Gong2014A}, LDA-based methods \cite{putthividhy2010topic} and neural network based methods \cite{liong2017deep,wang2017adversarial} also fall into this category. On the other hand, the cross-modal hashing methods mainly focus on the retrieval efficiency by mapping the items of different modalities into a common binary Hamming space.

Benefited from the strong ability of distribution modeling and discriminative representation learning, some recent cross-modal retrieval methods have collaborated with GAN models \cite{peng2017cm, huang2017mhtn, wang2017adversarial}. In this work, our method also follows the similar adversarial learning framework that uses the single-modal similarities to guide the cross-modal representation learning.
More recently, methods to explore the semantic relationships between unpaired items have been proposed. The ACMR \cite{wang2017adversarial} method proposes the triplet loss and the modality classifier for preserving the modality level semantic structures. The MHTN \cite{huang2017mhtn} is proposed to minimize the maximum mean discrepancy between modalities, which preserves more flexibility for the generator to project vectors into a new space. The difference between CMST and the previous work is that CMST can learn the item-level semantic relationships between unpaired items in an unsupervised way.


\section{CMST-based Cross-Modal Retrieval}
\subsection{Problem Formulation}
Without losing generality, we consider the images and texts pairs in this paper. Let $V = \left[ \mathbf{v}_1, \mathbf{v}_2, \dots, \mathbf{v}_n \right] \in \mathbb{R}^{d_v \times n}$ be a collection of image features, and $T = \left[ \mathbf{t}_1, \mathbf{t}_2, \dots, \mathbf{t}_n \right] \in \mathbb{R}^{d_t \times n}$ be the corresponding collection of text features, in which $\mathbf{v}_i$ and $\mathbf{t}_i$ form a pair, where $d_v$ and $d_t$ denote the dimension of the image features and the text features, respectively.
Each sample pair is assigned a semantic label vector denoted as $\mathbf{y_i} = \left[ y_{i1}, y_{i2}, \dots, y_{ic} \right] \in \mathbb{R}^c$, where $c$ indicates the semantic classes. If the $i$-th sample pair in $V$ and $T$ belongs to the semantic class $j$, $y_{ij} = 1$; otherwise, $y_{ij} = 0$. We denote the collection of semantic label vectors as $Y = \left[ \mathbf{y_1}, \mathbf{y_2}, \dots, \mathbf{y_n} \right] \in \mathbb{R}^{d_c \times n} $.
The goal of our proposed CMST method is to learn a common semantic space $S = \left[ \mathbf{s}_1, \mathbf{s}_2, \dots, \mathbf{s}_n \right] \in \mathbb{R}^{d_s \times n}$, where the features from different modalities can be directly compared in terms of semantic similarity, and $d_s$ denotes the dimension of the common semantic space.


\subsection{Learning the Single-modal Similarity}

The Siamese networks \cite{chopra2005learning} are adopted as similarity learning networks to learn the semantic relationships in single-modal representation subspace. Siamese networks can learn similarity metrics discriminatively and effectively. In addition, Siamese networks are also robust to data noises because they consider not only the label relationship but also the distance between the features of each sample pair. For simplicity, we only detail the network in image modality. Given two image features and their labels $(\mathbf{v}_i, \mathbf{y}_i)$ and $(\mathbf{v}_j, \mathbf{y}_j)$, let $u_{ij}=1$ if $\mathbf{y}_i = \mathbf{y}_j$, otherwise $u_{ij}=0$ . The loss function of our similarity learning network can be formulated as:
\begin{equation}
\small
\mathcal{L}_{sia}= \sum_{i,j} u_{ij} \cdot s(\mathbf{v}_i, \mathbf{v}_j) + (1 - u_{ij}) \cdot \max(C - s(\mathbf{v}_i, \mathbf{v}_j), 0).
\end{equation}
The similarity measurement in single-modal representation subspace is given as
\begin{equation}
\small
s(\mathbf{v}_i, \mathbf{v}_j) = || H_V(\mathbf{v}_i) - H_V(\mathbf{v}_j) ||_{2}^{2},
\end{equation}
where the $H_V$ denotes the image similarity learning network consists of feed-forward layers with ReLU activations. The text similarity learning network has the same structure as the image similarity learning network. The Siamese network is illustrated in the middle bottom part of Figure \ref{fig:2}.

\subsection{Transferring Learned Similarities to Common Representation Subspace}

\subsubsection{Value transferring}
A direct way of using the intra-modal similarity as a reference for cross-modal similarity learning is to use the value as the learning goal. In this case, we use $\mathbf{t}_j$ as the anchor sample, only the intra-modal similarity that is different from it (image-modality similarity) is used for clear explanation, but the training process contains both directions. Also, the range of the original features and the projected features are not limited, as the fully connected networks have the ability of scaling between the input and the expected output.
The typical form of similarity transferring loss based on value transferring is defined as
\begin{equation}
\small
\mathcal{L}_{val}=|s(\mathbf{v}_j, \mathbf{v}_j) - s(\mathbf{v}_j, \mathbf{t}_j)| + |s(\mathbf{v}_i, \mathbf{v}_j) - s(\mathbf{v}_i, \mathbf{t}_j)|,
\end{equation}
where the similarity between $\mathbf{v}_j$ and itself is set to value 1.

\subsubsection{Difference transferring}
Compared to the absolute value of intra-modal similarity itself, it is the relationship between cross-modal samples that we are really interested in. In other words, the relation of $(s(\mathbf{v}_i, \mathbf{t}_j) - s(\mathbf{v}_j, \mathbf{t}_j))$ can measure the difference of $\mathbf{v}_i$ to $\mathbf{t}_j$ and $\mathbf{v}_j$ to $\mathbf{t}_j$ in the common semantic space, where the difference is expected to be related to the difference of $\mathbf{v}_i$ and $\mathbf{v}_j$ in their original image space. By difference transferring, the range of the common semantic space and the original modality space can be decoupled. The learnt joint subspace can be more flexible to modality divergence.
The typical form of similarity transferring loss based on difference transferring is defined as
\begin{equation}
\small
\mathcal{L}_{diff} =|(s(\mathbf{v}_j, \mathbf{t}_j) - s(\mathbf{v}_i, \mathbf{t}_j)) - (s(\mathbf{v}_j, \mathbf{v}_j) - s(\mathbf{v}_i, \mathbf{v}_j))|,
\end{equation}
where we assume $s(\mathbf{v}_j, \mathbf{v}_j) = 1$.

\subsubsection{Product transferring} As transitivity exists in the nature of similarity measurement, similarity transitivity across modalities can be expected. Following this motivation, product transferring can be done by the multiplication on the similarity chain. We can say that the similarity of $\mathbf{v}_i$ and $\mathbf{t}_j$ is generated by the chain of $\mathbf{v}_i$ to $\mathbf{v}_j$ and then $\mathbf{v}_j$ to $\mathbf{t}_j$. In this case, the typical form of similarity transferring loss based on product transferring is defined as
\begin{equation}
\small
\mathcal{L}_{prod}=|s(\mathbf{v}_i, \mathbf{t}_j) - s(\mathbf{v}_i, \mathbf{v}_j)  s(\mathbf{v}_j, \mathbf{t}_j)|,
\end{equation}
where some implicit linear function is assumed to enable the direct arithmetical operation between inter-modal and intra-modal similarities. The linear function is absorbed into the function approximation ability of neural networks.

The proposed similarity transferring approaches provide a reference value or reference relationship for the unpaired items of two different modalities, so the loss function can be defined as the difference between the similarity calculated from learnt cross-modal features and the reference values. The similarity transferring task is applied together with other widely used tasks within a generative adversarial framework.

\begin{table*}[h]
\small
\centering
\caption{Comparison with existing cross-modal retrieval methods in aspect of mAP}
\label{tab:2}
\begin{minipage}{\textwidth}
\begin{center}
\begin{tabular}{lcccccccccc}
\toprule
\multirow{2}{1.8cm}{Category} & \multirow{2}{1cm}{Methods} & \multicolumn{3}{c}{Wikipedia} & \multicolumn{3}{c}{Pascal Sentences} & \multicolumn{3}{c}{NUS-WIDE-10k} \\
& & Img2txt & Txt2Img & Avg. & Img2txt & Txt2Img & Avg. & Img2txt & Txt2Img & Avg. \\ \hline
\multirow{4}{1.8cm}{Traditional methods}
 & CCA-3V \cite{Gong2014A} & 0.437 & 0.383 & 0.410 & 0.316 & 0.270 & 0.293 & - & - & - \\
& LCFS \cite{wang2013learning} & 0.455 & 0.398 & 0.427 & 0.442 & 0.357 & 0.400 & 0.383 & 0.346 & 0.365 \\
& JRL \cite{zhai2014learning} & 0.453 & 0.400 & 0.426 & 0.504 & 0.489 & 0.496 & 0.426 & 0.376 & 0.401 \\
& JFSSL \cite{Wang2016Joint} & 0.428 & 0.396 & 0.412 & - & - & -& - & - & -\\ \hline
\multirow{3}{1.8cm}{DNN-based methods}
& Corr-AE \cite{feng2014cross} & 0.402 & 0.395 & 0.398 & 0.489 & 0.444 & 0.467 & 0.366 & 0.417 & 0.392 \\
& DCML \cite{liong2017deep} & 0.554 & 0.538 & 0.546 & - & - & - & 0.385 & 0.405 & 0.395 \\
& CMDN \cite{peng2016cross} & 0.488 & 0.427 & 0.458 & 0.534 & 0.534 & 0.534 & 0.492 & 0.515 & 0.504 \\ \hline
\multirow{4}{1.8cm}{GAN-based methods}
& CM-GAN \cite{peng2017cm} & 0.521 & 0.466 & 0.494 & 0.603 & 0.604 & 0.604 & - & - & -\\
& MHTN \cite{huang2017mhtn} & 0.514 & 0.444 & 0.479 & 0.496 & 0.500 & 0.498 & 0.520 & 0.534 & 0.527\\
& ACMR \cite{wang2017adversarial} & 0.619 & 0.489 & 0.554 & 0.535 & 0.543 & 0.539 & 0.544 & 0.538 & 0.541\\ 
& CMST (Ours) & \textbf{0.632} & \textbf{0.505} & \textbf{0.569} &\textbf{0.621} & \textbf{0.586} & \textbf{0.604} & \textbf{0.628} & \textbf{0.562} & \textbf{0.595} \\
\bottomrule
\end{tabular}
\end{center}
\end{minipage}
\end{table*}

\subsection{Total Loss for Training}
Follow the common practice of cross-modal retrieval, the proposed CMST introduces the adversarial learning and classification task for learning a better semantic structure in common representation subspace.

Let $\mathcal{L}_{sim}$ denote the similarity transferring task, which is chosen as the one from $\mathcal{L}_{val}$, $\mathcal{L}_{diff}$ and $\mathcal{L}_{prod}$. For classification task, the cross entropy loss is used in CMST and denoted as $\mathcal{L}_{lab}$. For adversarial learning, the discriminator composed of $3$ feed-forward layers takes the generated representation as input. The output is the prediction of which modality the input comes from, using the sigmoid activation. Let $\mathcal{L}_{V}$ denotes the cross entropy loss of predicting the image input, and $\mathcal{L}_{T}$ denotes the cross entropy loss of predicting the text input. The total loss for generator and the discriminator is given as
\begin{equation}
\small
\mathcal{L}_{G} = \mathcal{L}_{lab} + \mathcal{L}_{sim} + \mathcal{L}_{V} - \mathcal{L}_{T},
\end{equation}
\begin{equation}
\small
\mathcal{L}_{D} = - \mathcal{L}_{V} + \mathcal{L}_{T}
\end{equation}

\section{Experiments}
\subsection{Experimental Setup}
Three widely used datasets for cross-modal retrieval are used in the experiments, including Wikipedia dataset \cite{pereira2014role}, NUS-WIDE-10k dataset \cite{chua2009nus} and Pascal Sentence dataset \cite{rashtchian2010collecting}.
We follow the dataset partition method as  \cite{wang2017adversarial} for fair comparison. Image features are taken from the fc7 layer of a pre-trained VGGNet-19 model while text features are computed by the classic Bag-of-Words features with tf-idf weighting.
 The image features for all the datasets are of 4096 dimensions, while the text BoW feature is 5000 dimensions for the Wikipedia dataset, 1000 dimensions for the NUS-WIDE-10k and Pascal datasets.

For the class-based retrieval task, performance is measured in terms of the mean average precision (mAP), the measurement is applied on both directions, i.e. Img2txt and Txt2img. The larger the mAP value is, the better the performance becomes.
For the pair-based retrieval task, we accept only the ground-truth paired sample of the input query as correct retrieval result. The performance is evaluated by top-$k$ accuracy, indicating the times that the correct retrieval result appears within the top-$k$ retrieved results over the test set.

\subsection{Results and Analysis}
The CMST is compared with three classes of cross-modal retrieval methods, namely traditional methods, DNN-based methods and GAN-based methods, as shown in Table \ref{tab:2}. The results of CMST is based on difference transferring.

For the results shown in the table, the number of the intra-class samples with the query are counted in top-50 retrieved documents. Results show that our CMST method outperforms the counterparts on all three datasets for cross-modal retrieval tasks. On the Pascal dataset, our method improves the best competitor ACMR by 16.1\% and 7.9\% in image to text and text to image retrieval tasks, respectively. On NUS-WIDE-10k and Wikipedia dataset, the proposed CMST method achieves a relatively small but solid improvement compared to the state-of-the-art performance. The underlying cause why the method improves a lot on Pascal dataset but not as much on the other two datasets is that the Pascal dataset contains more semantic classes than the other two datasets. On Wikipedia and NUS-WIDE-10k, although the intra-class and inter-class labels are coarse supervision for cross-modal similarity learning, the small number of total classes makes it easy to separate between classes and arrange the joint semantic distribution. For the datasets with more classes, such as Pascal Sentences, the finer similarity structure benefits the generation of common semantic space as we expected.


\begin{table*}[h]
\small
\centering
\caption{Comparison with ACMR method on Pascal Sentences dataset in aspect of top-k acc}
\label{tab:3}
\begin{minipage}{\textwidth}
\begin{center}
\begin{tabular}{C{1.7cm}C{0.8cm}C{0.8cm}C{0.8cm}C{0.8cm}C{0.8cm}C{0.7cm}C{0.8cm}C{0.8cm}C{0.8cm}C{0.8cm}C{0.8cm}C{0.8cm}}
\toprule
\multirow{2}{1.0cm}{Methods} & \multicolumn{3}{c}{$k=1$} & \multicolumn{3}{c}{$k=5$} &  \multicolumn{3}{c}{$k=10$} & \multicolumn{3}{c}{$k=50$} \\
 & Img2txt & Txt2Img & Avg. & Img2txt & Txt2Img & Avg. & Img2txt & Txt2Img & Avg. & Img2txt & Txt2Img & Avg.\\ \hline
ACMR & 0.140 & 0.145 & 0.143 & 0.445 & 0.400 & 0.423 & 0.675 & 0.665 & 0.670 & 0.910 & 0.915 & 0.913 \\
CMST(Ours) & \textbf{0.210} & \textbf{0.200} & \textbf{0.205} & \textbf{0.455} & \textbf{0.550} & \textbf{0.503} & \textbf{0.725} & \textbf{0.750} & \textbf{0.738} & \textbf{0.990} & \textbf{0.965} & \textbf{0.978} \\
\bottomrule
\end{tabular}
\end{center}
\end{minipage}
\end{table*}

Table \ref{tab:3} shows the experimental results in the pair-based retrieval task in aspect of top-$k$ accuracy on Pascal dataset, where different $k$ values are tested to examine the method's performance. For smaller values of $k$ ($k$=1,5,10), our proposed method outperforms the ACMR method by 43.4\%, 18.9\% and 10.1\% for the average measurement on two retrieval directions. The results indicate that our proposed CMST method works effectively on improving the rank of the most related retrieval results for the given query.

\subsection{Further Analysis on CMST}
In this subsection, several experiments are conducted to investigate the effectiveness of the important components of CMST method.

\begin{table}[h]
\small
\centering
\caption{Similarity transferring with different intra-modal similarity measurements}
\label{tab:4}
\begin{minipage}{\columnwidth}
\begin{center}
\begin{tabular}{lcccc}
\toprule
\multirow{2}{1cm}{Methods} & \multicolumn{2}{c}{mAP} & \multicolumn{2}{c}{top-1 acc} \\
 & Img2txt & Txt2Img & Img2txt & Txt2Img \\ \hline
Cosine & 0.509 & 0.500 & 0.150 & 0.140 \\
Euclidean & 0.563 & 0.554 & 0.165 & 0.125 \\
Siamese Network & \textbf{0.621} & \textbf{0.586} & \textbf{0.210} & \textbf{0.200} \\
\bottomrule
\end{tabular}
\end{center}
\end{minipage}
\end{table}

The intra-modal similarity learnt by the Siamese networks plays an important role in the subsequent similarity transferring procedure. To examine its effectiveness, two traditional similarity measurements are employed as the source of intra-modal similarity for comparison. Table \ref{tab:4} illustrates the performance of traditional similarity metric and the learnt Siamese similarity metric, showing that the Siamese network achieves the best performance among the three similarity measurements.

\begin{table}[h]
\small
\centering
\caption{Effects of the similarity transferring approaches}
\label{tab:5}
\begin{minipage}{\columnwidth}
\begin{center}
\begin{tabular}{lcccc}
\toprule
\multirow{2}{1cm}{Methods} & \multicolumn{2}{c}{mAP} & \multicolumn{2}{c}{top-1 acc}  \\
 & Img2txt & Txt2Img & Img2txt & Txt2Img \\ \hline
Value & 0.600 & 0.564 & 0.145 & 0.155 \\
Difference & \textbf{0.621} & \textbf{0.586} & \textbf{0.210} & \textbf{0.200} \\
Product  & 0.510 & 0.514 & 0.120 & 0.140 \\
No transfer  & 0.485 & 0.508 & 0.080 & 0.105 \\
\bottomrule
\end{tabular}
\end{center}
\end{minipage}
\end{table}

In order to compare the different methods of similarity transferring, we examine the three approaches on Pascal dataset. The results in the last row in Table \ref{tab:5} with no transferring indicates that the similarity transferring task is not included in the training. Overall, we can see that all the similarity transferring approaches are effective for noticeable performance improvement with similarity transferring training. The best similarity transferring approach contributes an improvement of 21.6\% compared to the CMST without similarity transferring. The difference transferring approach performs better than the value transferring method because it keeps the comparative relationship between samples instead of assigning a value to the similarity between samples. The generator is guaranteed with more flexibility for the feature projection by difference transferring method. The product transferring seems to perform badly, as the distance measurement used in the experiment is the Euclidean distance and there is no explicit limitations or normalization methods added to the distance calculation. The product of the Euclidean distances between high dimensional feature vectors varies sharply on minor changes.

\begin{table}[h]
\small
\centering
\caption{Effects of different training strategies.}
\label{tab:6}
\begin{minipage}{\columnwidth}
\begin{center}
\begin{tabular}{lcccc}
\toprule
\multirow{2}{1cm}{Strategies} & \multicolumn{2}{c}{mAP} & \multicolumn{2}{c}{top-1 acc}  \\
 & Img2txt & Txt2Img & Img2txt & Txt2Img \\ \hline
Two-stage & \textbf{0.621} & \textbf{0.586} & \textbf{0.210} & \textbf{0.200} \\
Fine-tuning & 0.596 & 0.571 & 0.180 & 0.180 \\
End-to-end  & 0.585 & 0.561 & 0.140 & 0.165 \\
\bottomrule
\end{tabular}
\end{center}
\end{minipage}
\end{table}

A two-stage training strategy is employed by firstly learning the single-modal similarity metric and then transferring the single-modal similarity into cross-modal semantic space. Note that all previous results are based on the two-stage strategy. On the other hand, fine-tuning and end-to-end training strategies are also available for our proposed CMST method. To examine the influence of different training strategies, we conduct an additional experiment by using different training method and compare the testing results at the same epoch (100). In the two-stage training and fine-tuning training, the Siamese networks are trained for 50 epoches in advance. For two-stage training, the parameters of Siamese networks are fixed after the 50th epoch, while for fine-tuning, the parameters are still learnable with a low learning rate (0.0001). Table \ref{tab:6} shows the effect of different training method, and we can draw from the table that two-stage training yields the best performance. Fine-tuning does not provide additional performance improvement. End-to-end training has negative influence on cross-modal similarity learning because the single-modal similarity learning should not be heavily affected by cross-modal information.

\section{Conclusions}
In this paper, a novel cross-modal retrieval method named CMST is proposed. The proposed method efficiently learns similarity metrics in each single modality space by the intra-modal similarity learning networks and then guide the cross-modal similarity learning with the learnt single modal similarity metric. Our proposed similarity transferring approaches successfully transfer finer similarity structure captured in single modal space to cross-modal space. Experiments demonstrate that the CMST method outperforms state-of-the-art cross-modal retrieval methods in both class-based and pair-based retrieval tasks.
\section{Acknowledgments}
This work was supported by National Key R\&D Program of China (2018YFB0505400), in part by Tsinghua-Kuaishou Institute of Future Media Data, and we also thank Mr. Tao Li and Ms. Junhui Wu, Kuaishou’s R\&D engineers, for their assistance in the research.

\bibliographystyle{IEEEbib}
\bibliography{CMST}

\begin{thebibliography}{10}

\bibitem{liong2017deep}
Venice~Erin Liong, Jiwen Lu, Yap-Peng Tan, and Jie Zhou,
\newblock ``Deep coupled metric learning for cross-modal matching,''
\newblock {\em IEEE Transactions on Multimedia}, vol. 19, no. 6, pp.
  1234--1244, 2017.

\bibitem{wang2017adversarial}
Bokun Wang, Yang Yang, Xing Xu, Alan Hanjalic, and Heng~Tao Shen,
\newblock ``Adversarial cross-modal retrieval,''
\newblock in {\em ACM MM}. ACM, 2017, pp. 154--162.

\bibitem{mignon2012cmml}
Alexis Mignon and Fr{\'e}d{\'e}ric Jurie,
\newblock ``Cmml: A new metric learning approach for cross modal matching,''
\newblock in {\em ACCV}, 2012.

\bibitem{song2018binary}
Jingkuan Song, Tao He, Lianli Gao, Xing Xu, Alan Hanjalic, and Heng~Tao Shen,
\newblock ``Binary adversarial networks for image retrieval,''
\newblock in {\em AAAI}, 2018.

\bibitem{xu2017learning}
Xing Xu, Fumin Shen, Yang Yang, Heng~Tao Shen, and Xuelong Li,
\newblock ``Learning discriminative binary codes for large-scale cross-modal
  retrieval,''
\newblock {\em TIP}, vol. 26, no. 5, pp. 2494--2507, 2017.

\bibitem{andrew2013deep}
Galen Andrew, Raman Arora, Jeff Bilmes, and Karen Livescu,
\newblock ``Deep canonical correlation analysis,''
\newblock in {\em ICML}, 2013, pp. 1247--1255.

\bibitem{Gong2014A}
Yunchao Gong, Qifa Ke, Michael Isard, and Svetlana Lazebnik,
\newblock ``A multi-view embedding space for modeling internet images, tags,
  and their semantics,''
\newblock {\em International Journal of Computer Vision}, vol. 106, no. 2, pp.
  210--233, 2014.

\bibitem{putthividhy2010topic}
Duangmanee Putthividhy, Hagai~T Attias, and Srikantan~S Nagarajan,
\newblock ``Topic regression multi-modal latent dirichlet allocation for image
  annotation,''
\newblock in {\em CCVPR}. IEEE, 2010, pp. 3408--3415.

\bibitem{peng2017cm}
Yuxin Peng, Jinwei Qi, and Yuxin Yuan,
\newblock ``Cm-gans: Cross-modal generative adversarial networks for common
  representation learning,''
\newblock {\em arXiv preprint arXiv:1710.05106}, 2017.

\bibitem{huang2017mhtn}
Xin Huang, Yuxin Peng, and Mingkuan Yuan,
\newblock ``Mhtn: Modal-adversarial hybrid transfer network for cross-modal
  retrieval,''
\newblock {\em arXiv preprint arXiv:1708.04308}, 2017.

\bibitem{chopra2005learning}
Sumit Chopra, Raia Hadsell, and Yann LeCun,
\newblock ``Learning a similarity metric discriminatively, with application to
  face verification,''
\newblock in {\em CVPR}. IEEE, 2005, pp. 539--546.

\bibitem{wang2013learning}
Kaiye Wang, Ran He, Wei Wang, Liang Wang, and Tieniu Tan,
\newblock ``Learning coupled feature spaces for cross-modal matching,''
\newblock in {\em ICCV}. IEEE, 2013, pp. 2088--2095.

\bibitem{zhai2014learning}
Xiaohua Zhai, Yuxin Peng, and Jianguo Xiao,
\newblock ``Learning cross-media joint representation with sparse and
  semisupervised regularization,''
\newblock {\em IEEE Transactions on Circuits and Systems for Video Technology},
  vol. 24, no. 6, pp. 965--978, 2014.

\bibitem{Wang2016Joint}
Kaiye Wang, Ran He, Liang Wang, Wei Wang, and Tieniu Tan,
\newblock ``Joint feature selection and subspace learning for cross-modal
  retrieval,''
\newblock {\em TPAMI}, vol. 38, no. 10, pp. 2010--2023, 2016.

\bibitem{feng2014cross}
Fangxiang Feng, Xiaojie Wang, and Ruifan Li,
\newblock ``Cross-modal retrieval with correspondence autoencoder,''
\newblock in {\em ACM MM}. ACM, 2014, pp. 7--16.

\bibitem{peng2016cross}
Yuxin Peng, Xin Huang, and Jinwei Qi,
\newblock ``Cross-media shared representation by hierarchical learning with
  multiple deep networks,''
\newblock in {\em IJCAI}, 2016, pp. 3846--3853.

\bibitem{pereira2014role}
Jose~Costa Pereira, Emanuele Coviello, Gabriel Doyle, Nikhil Rasiwasia, Gert~RG
  Lanckriet, Roger Levy, and Nuno Vasconcelos,
\newblock ``On the role of correlation and abstraction in cross-modal
  multimedia retrieval,''
\newblock {\em TPAMI}, vol. 36, no. 6, pp. 521--535, 2014.

\bibitem{chua2009nus}
Tat-Seng Chua, Jinhui Tang, Richang Hong, Haojie Li, Zhiping Luo, and Yantao
  Zheng,
\newblock ``Nus-wide: a real-world web image database from national university
  of singapore,''
\newblock in {\em Proceedings of the ACM International Conference on Image and
  Video Retrieval}. ACM, 2009.

\bibitem{rashtchian2010collecting}
Cyrus Rashtchian, Peter Young, Micah Hodosh, and Julia Hockenmaier,
\newblock ``Collecting image annotations using amazon's mechanical turk,''
\newblock in {\em NAACL HLT}. ACL, 2010, pp. 139--147.

\end{thebibliography}

\end{document}